\documentclass[sigconf,article]{acmart}
\usepackage{graphicx}
\usepackage{tcolorbox}
\usepackage{color}
\usepackage{subcaption}

\renewcommand\footnotetextcopyrightpermission[1]{} 
\begin{document}

\title{LifeIR at the NTCIR-18 Lifelog-6 Task}

\author{Jiahan Chen}
\affiliation{
    \institution{Key Laboratory of Network Data Science and Technology, Institute of Computing Technology, Chinese Academy of Sciences}
    {State Key Laboratory of AI Safety}
    {University of Chinese Academy of Sciences}
    \country{China}
}
\email{chenjiahan24s@ict.ac.cn}

\author{Da Li}
\affiliation{
    \institution{Key Laboratory of Network Data Science and Technology, Institute of Computing Technology, Chinese Academy of Sciences}
    {State Key Laboratory of AI Safety}
    {University of Chinese Academy of Sciences}
    \country{China}
}
\email{lida21s@ict.ac.cn}
\author{Keping Bi}
\authornote{Corresponding author.}
\affiliation{
    \institution{Key Laboratory of Network Data Science and Technology, Institute of Computing Technology, Chinese Academy of Sciences}
    {State Key Laboratory of AI Safety}
    {University of Chinese Academy of Sciences}
    \country{China}
}

\email{bikeping@ict.ac.cn}

\begin{abstract}
In recent years, sharing lifelogs recorded through wearable devices such as sports watches and GoPros, has gained significant popularity. Lifelogs involve various types of information, including images, videos, and GPS data, revealing users' lifestyles, dietary patterns, and physical activities. The Lifelog Semantic Access Task(LSAT) in the NTCIR-18 Lifelog-6 Challenge focuses on retrieving relevant images from a large scale of users' lifelogs based on textual queries describing an action or event. It serves users' need to find images about a scenario in the historical moments of their lifelogs. We propose a multi-stage pipeline for this task of searching images with texts, addressing various challenges in lifelog retrieval. Our pipeline includes: filtering blurred images, rewriting queries to make intents clearer, extending the candidate set based on events to include images with temporal connections, and reranking results using a multimodal large language model(MLLM) with stronger relevance judgment capabilities. The evaluation results of our submissions have shown the effectiveness of each stage and the entire pipeline.
\end{abstract}


\maketitle
\pagestyle{plain} 

\section*{Team Name} 
LifeIR

\section*{Subtasks}
Lifelog Semantic Access Subtask (LSAT) - Automatic

\section{Introduction}
Lifelogging is a behavior of users recording their daily activities through wearable devices, smartphones, or ambient sensors~\cite{INR-033}. It captures multimodal data encompassing visual content (e.g., photos and videos), spatiotemporal information (e.g., GPS trajectories and timestamps), physiological signals (e.g., heart rate and step counts), and behavioral interactions (e.g., app usage logs)~\cite{ribeiro2022lifelog}. With the advancement of the Internet of Things (IoT), lifelogging has emerged as a critical research area in digital memory, personalized healthcare, and human behavior analysis. Lifelogging holds significant research value as it enables continuous, objective recording and analysis of human behavior and physiological states in real-world settings. The large scale of lifelog data stimulates the demand for retrieval of specific life episodes, facilitates deeper comprehension of human cognitive patterns, and supports the derivation of individualized behavioral inferences. With the landscape of lifelogging research initiatives, the NTCIR workshop has formulated the Lifelog Semantic Access Subtask (LSAT), formally defined as a non-interactive (fully automated) known-item search task. In this subtask, researchers are required to 
retrieve historical images about a target scene from the recorded videos in their lifelogs given textual queries. It is a know-item search task similar to email search and has become increasingly popular as smartphones and smartwatches accumulate large amounts of data. As the core competition component of the NTCIR-Lifelog challenge series for multiple consecutive years, LSAT has effectively stimulated academic efforts to develop optimal solutions for this crucial scientific problem of lifelog moment retrieval.

Compared to textual data, lifelogs exhibit three characteristics that feature the differences between lifelog retrieval and text retrieval. They are: (1) \textbf{Data Heterogeneity}: Data are collected from various sensors (e.g., images, text, sensor signals) that vary in structure and semantics~\cite{jirkovsky2016understanding}. This variability necessitates the development of cross-modal alignment and fusion techniques. (2) \textbf{Temporal Connection}: Life events demonstrate significant temporal dependencies~\cite{sreekumar2014geometry}, such as: ``drinking
coffee before a morning meeting''. These temporal dependencies pose substantial challenges for conventional keyword-based retrieval systems and visual recognition methods~\cite{tian2021probabilistic}. (3) \textbf{Query Intent Specificity}: In lifelong retrieval, users search an image or episode in their lifelogs that they know existing. Their queries mainly describe the scene or episode. The specific details in the queries are important for differentiating the targeted images. An effective lifelog retrieval system should take these characteristics into account. 

Our LifeIR team participated in the Lifelog Semantic Access Task (LSAT) of the NTCIR-18 Lifelog-6 challenge~\cite{ntcir18-lifelog6}. In this paper, we concentrate on the LSAT subtask performed in an automated manner. We propose some steps to improve the performance of the LifeIR retrieval algorithm. Our method is based on the image-text embedding representation derived from the CLIP~\cite{radford2021learningtransferablevisualmodels} and improves the retrieval results by adding modules. Building upon the three aforementioned characteristics of lifelogs, we propose our solution for the LSAT task with the following key components:
(1) \textbf{Data Heterogeneity}: To align visual and textual modalities, we employ the CLIP model during the retrieval stage and introduce a Multimodal Large Language Model (MLLM) in the re-ranking phase. (2) \textbf{Temporal Connection}: Firstly, we construct the candidate event set by calculating the similarity between two images adjacent to each other on the same day. Subsequently, we leverage the timestamp-based naming feature of images to expand the candidate images after the first-stage retrieval.  (3) \textbf{Query Intent Specificity}: We reformulate the initial query for each topic to preserve critical details from the original query while controlling its length. This ensures CLIP’s sensitivity to the reformulated queries, thereby improving initial retrieval accuracy. Our experimental results demonstrate that: (1) Query reformulation is effective, enabling CLIP to retrieve images corresponding to specific moments described in the query. (2) Both temporal and similarity-based methods significantly enhance the candidate set after initial retrieval. (3) Leveraging MLLM’s robust cross-modal alignment capability for re-ranking effectively filters out pseudo-relevant images, ultimately improving retrieval precision.

The rest of the paper is structured as follows: Section 2 briefly describes the related work. Then, in Section 3, we introduce our method, covering the core aspects of the approach such as filtering blurred images, rewriting queries, the multi-round extension of the candidate set based on events, and so on. We present and discuss the results of the experiment in Section 4. Finally, in Section 5, we conclude our paper and discuss possible improvements of our method.

\section{Related Work}
In recent years, there has been increasing academic interest in lifelogging, particularly exemplified by benchmark competitions such as the Lifelog Search Challenge (LSC) and NTCIR-lifelog task. These initiatives have led significant advancements in multimodal retrieval and temporal reasoning.

\textbf{Existing works on lifelog retrieval.} In LSC23~\cite{10.1145/3591106.3592304}, Memento 3.0~\cite{10.1145/3592573.3593103} implements a cross-model fusion technique, where image-query similarity scores from two independent CLIP embeddings are combined to improve retrieval ranking. Voxento 4.0~\cite{10.1145/3592573.3593097} introduces a voice-enabled lifelog retrieval system, improving accessibility through multimodal retrieval. It employs CLIP for visual feature extraction and provides the option of using a text-based retrieval to complement the voice-based method. LifeXplore~\cite{10.1145/3592573.3593105} introduces a concept-based image retrieval system powered by ensemble modeling. CLIP bridges textual and visual modalities, and then CRAFT~\cite{DBLP:journals/corr/abs-1904-01941} extracts embedded text. Finally, YOLOv7~\cite{wang2022yolov7trainablebagoffreebiessets} localizes objects, and EfficientNet B2 abstracts high-level semantics. This pipeline automates descriptive tag generation to optimize search accuracy. In the previous edition of Lifelog at NTCIR-17~\cite{zhou2023overview}, MemoriEase~\cite{tran2024memoriease} leveraged the power of the BLIP-2~\cite{li2023blip} model to retrieve lifelog images efficiently in response to text queries. It incorporated ChatGPT for both preprocessing and post-processing tasks. LifeInsight~\cite{nguyen2023lifeinsight} first applied Semantic Role Labeling (SRL) to decompose queries into structured entities while enhancing metadata representations. Subsequently, they employed advanced LLMs to temporally ground events and dynamically produce context-aware prompts. DCU-Memento integrated an ensemble of CLIP variants, incorporating both standard OpenAI implementations and data-enhanced OpenCLIP architectures. The system also implemented a hierarchical retrieval pipeline: primary ranking through visual feature matching followed by context-aware filtering for result refinement. These systems collectively demonstrate critical technological paradigms for lifelog retrieval systems, while simultaneously delineating both prevailing research trajectories and persistent challenges in the field. 

\textbf{Evolution of image-text retrieval models.} Modern image retrieval systems have evolved from early bag-of-words models to sophisticated cross-modal architectures. Foundation models like CLIP (Contrastive Language-Image Pretraining) revolutionized the field by learning joint embedding spaces through contrastive loss on 400M image-text pairs. Subsequent variants (JINA-CLIP) extended this paradigm to multilingual domains with improved fine-grained alignment. The emergence of Multimodal LLMs introduced generative retrieval capabilities, enabling query-aware image representations through cross-attention mechanisms. Multimodal Large Language Models (MLLMs) are models that extend traditional LLMs by integrating and processing multiple data modalities (e.g., vision, language) to achieve cross-modal understanding, reasoning, and generation capabilities. Early works like VisualBERT~\cite{li2019visualbert} established basic cross-modal alignment architectures, while CLIP advanced contrastive vision-language pretraining. The emergence of powerful LLMs enabled systems like Flamingo~\cite{alayrac2022flamingo} and BLIP-2~\cite{li2023blip} that integrated visual encoders with frozen language models through innovative adapter designs. Recent MLLMs demonstrate remarkable versatility, with GPT-4V showcasing production-grade multimodal understanding and open-source alternatives like LLaVA~\cite{liu2023visual}, Qwen~\cite{yang2024qwen2technicalreport} and MiniGPT-4~\cite{zhu2023minigpt} achieving competitive performance through improved visual feature alignment. MLLMs are able to optimize the intermediate process of multimodal retrieval and improve the effectiveness of the retrieval system. MLLMs can simulate user feedback. In some time-insensitive scenarios, MLLMs can even be used directly as a backbone of the retriever.

\section{Method}
\begin{figure*}[t!]
	\centering
	\includegraphics[width=0.95\textwidth]{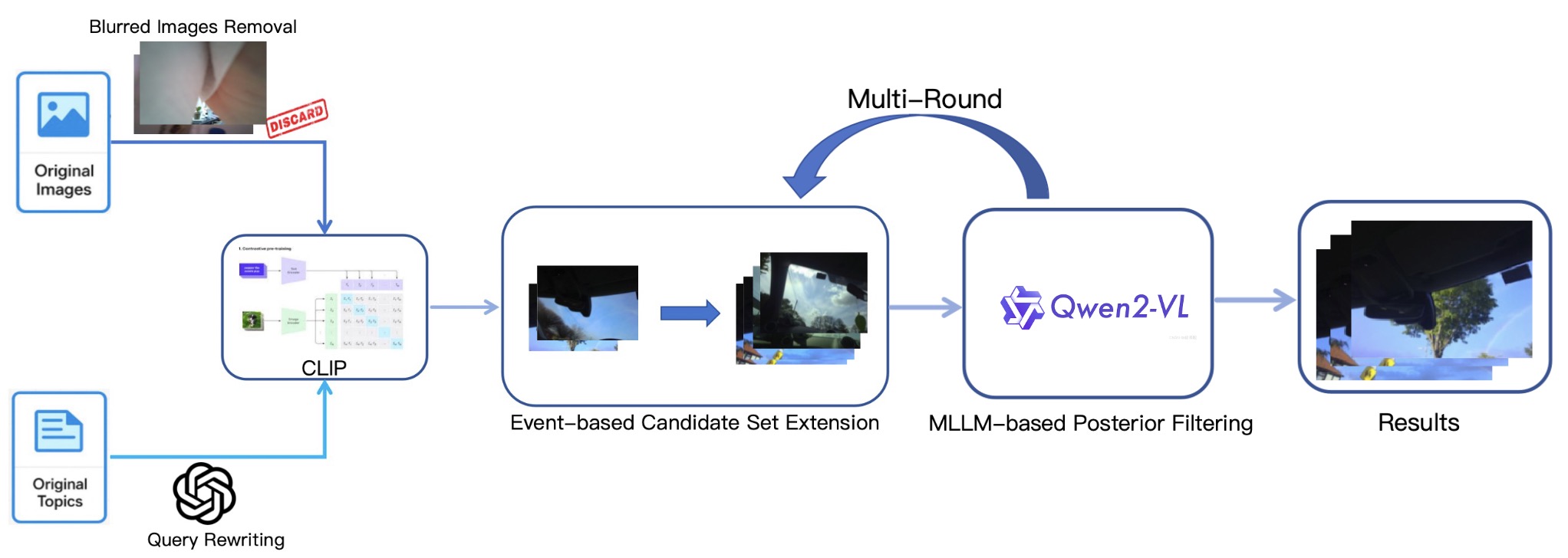} %
	\caption{Overview of Our Method.}
	\label{fig:example}
\end{figure*}

In this section, we present our method to construct the automatic retrieval method to address the Lifelog-6 subtask -- LSAT. Unlike those interactive lifelog retrieval systems, our automatic retrieval approach focuses on optimizing the results by improving the retrieval processes, rather than handing over the results to human to iterate to achieve better evaluation results. 
Figure \ref{fig:example} shows an overview of our method. Our retrieval method is divided into five main modules: (1) Data Cleaning, (2) CLIP-based Retrieval, (3) Query Rewriting, (4) Event-based Candidate Set Expansion, and (5) Posterior Filtering using MLLM. Our filtering and processing of the dataset is described in Section~\ref{data_cleaning}. Section~\ref{clip_based_retrieval} describes how we use the CLIP for preliminary retrieval. Section~\ref{query_rewrite} introduces the process of rewriting queries with LLM, and Section~\ref{event_extension} describes the validity of the event-based extension of the candidate set for our method. Finally, Section~\ref{mllm_filter} shows how to filter experimental results using MLLM.

\subsection{Data Cleaning}\label{data_cleaning} 
\textbf{Data Composition.} The dataset provided over 725k images from January 2019 to June 2020, spanning 18 months. Accompanying the lifelog images are metadata and visual concepts. The metadata encompasses information such as time, physical activities, biometrics, and locations when the image was taken. 
The visual concepts include detected scenes and concepts for each image.
All images in the dataset were obtained from one GoPro wearer. 

There are a total of 26 topics in the NTCIR-18 Lifelog-6, consisting of 13 ad-hoc topics and 13 known-item topics. All topics are made up of Title, Description, and Narrative. Title is a concise description of the topic. Description provides additional details and context to enhance understanding. Narrative is to topic and retrieve the corresponding image requirements. A simple case is shown in Table~\ref{tab:case_show}.
\begin{table}[h]
\renewcommand{\arraystretch}{1.2} 
\caption{A Query Topic Overview.}
\label{tab:case_show}
\begin{tabular}{lp{6cm}c}
\toprule
\textbf{Title} &Photographing meals. \\
\textbf{Description} &Find all the times I take a photo of my meal.\\
\textbf{Narrative} &Each instance should involve photographing a meal or plate of food, not just seeing or eating it. Taking a photo with a camera or phone are required for any eating instance to be relevant.\\ 
\bottomrule
\end{tabular}
\end{table}

\textbf{Image Filter.} Images are collected randomly, and due to variations in lighting and timing, the occurrence of blurred images is unavoidable. Given the large number of original images collected in the dataset, the initial step to remove these blurred images from the dataset. We use the edge weight summation method~\cite{canny1986computational} to filter images. The edge weight summation method provides an effective solution for detecting blurry images by quantifying the sharpness of edges within an image. This approach is divided into three steps: (1) edge feature extraction using gradient operators, (2) normalized edge density computation, and (3) adaptive thresholding for quality classification. We randomly sample the images in the dataset that are manually judged to be blurry and take their mean value as the filtering threshold. After filtering out more than 120k images, we got the new dataset. As for metadata, we analyze and retain fields that might be useful for retrieval (such as location) and use them later in the retrieval process.

\subsection{CLIP-based Retrieval}\label{clip_based_retrieval}
Given CLIP's demonstrated efficacy in prior visual-language retrieval competitions~\cite{nguyen2023lifeinsight}, our method adopts its architecture to independently encode queries and images. CLIP is a vision-language model that learns joint embeddings for images and text through contrastive training on large-scale image-text pairs. Its key strength lies in projecting both modalities into a shared latent space, enabling direct similarity computation between visual and textual representations without task-specific fine-tuning. We employ CLIP as our foundational retrieval model to address data heterogeneity by aligning visual and textual modalities into a unified embedding space. Given a query $Q$ and an image $I$, their embeddings $\mathbf{t}$ and $\mathbf{v}$ are computed via CLIP's visual and textual encoders, respectively. The relevance score $sim(Q,I)$ is measured by cosine similarity:
\begin{equation}
sim(Q,I) = \frac{\mathbf{t} \cdot \mathbf{v}}{\|\mathbf{t}\| \|\mathbf{v}\|}
\end{equation}
where $\mathbf{t} = \text{CLIP}_{\text{text}}(Q)$ and $\mathbf{v} = \text{CLIP}_{\text{vis}}(I)$ denote the L2-normalized embeddings.

\subsection{Query Rewriting}\label{query_rewrite}
The explicit expression of specific details in user queries is critical for effective lifelog retrieval. Lifelog queries typically target precise moments in personal history. Such queries must retain fine-grained descriptors - including temporal markers, spatial context, and activity details - to uniquely identify the sought-after memory among visually similar episodes. As shown in Table~\ref{tab:case_show}, the brief title of the query lacks the detailed information needed for relevance judgment. The description of the relevant image is partially available in the description and narrative. Using one of them alone as a query is missing information. The information provided by all three parts needs to be combined to get a complete query.

However, the model's pretraining paradigm on image-caption pairs introduces inherent limitations: its contrastive learning objective favors short text fragments.~\cite{chen2025lrsclipvisionlanguagefoundationmodel,zhang2024long}. To address this, we leverage the advanced natural language understanding capabilities of large language models (LLMs) to perform semantic query reformulation based on the given topics. We use ChatGPT for query rewriting. To ensure that the rewritten queries retain the original information of the topics and still describe the retrieval event from a first-person perspective, the prompt we built is shown in Figure~\ref{fig:query_rewriting}.
. To maximize CLIP's ability to understand the text, we limit the length of a rewritten query to 30 words.
\begin{figure}[h]
\begin{tcolorbox}[colback=white, colframe=black, sharp corners=southwest, boxrule=0.3pt]
\textbf{Rewriting Prompt:} \\
\textsc{Round 1} \\
I want to find images that meet the text requirements below. Please summarize in concise language the elements that must be present in the image. Unnecessary information and explanatory content cannot be output. \\
\textbf{[Query]}: Photographing meals.,  \\ 
\textbf{[Requirements]}: Find all the times I take a photo of my meal. Each instance should involve photographing a meal or plate of food, not just seeing or eating it. Taking a photo with a camera or phone are required for any eating instance to be relevant.\\
\textbf{[Output]}: \textit{Images must show instances of meals being photographed with a camera or phone, not just seen or eaten.} \\
\hrule
\vspace{1em}

\textsc{Round 2} \\
Use a first-person perspective to simply describe what an image containing the following contents would look like if recorded by a GoPro. If not mentioned, do not include anyone other than the GoPro wearer in the image. \\
There are some other requirements: \\
1. Do not use any qualifying or beautifying words. \\
2. Do not exceed 30 words.\\
3. Do not output information that is not in the content.\\
4. If there is no background description in the content, do not add background information such as weather.\\
5. Do not output GoPro. \\
\textbf{[Requirements]}: \textit{Images must show instances of meals being photographed with a camera or phone, not just seen or eaten.}  \\ 
\textbf{{[Output]}}: \textit{I'm taking a photo of my meal, aiming my camera or phone at it, and there it appears on the screen, capturing the moment before eating.}
\end{tcolorbox}
\caption{A Case of Rewriting Rewriting. The \textit{italicized text} is the output of the LLM during the rewrite process.}
\label{fig:query_rewriting}
\end{figure}

\subsection{Event-based Candidate Set Expansion}\label{event_extension}
\textbf{Temporal connection.} In the NTCIR-18 Lifelog-6, Queries are often aimed at searching for one or more semantic events. Semantic events are moments in everyday life that demonstrate a specific activity. They are usually temporally continuous, depending on the duration of the activity. The temporal continuity inherent in lifelog data provides critical insight for identifying relevant images. Unlike isolated images, lifelog data is time coherent — where temporally adjacent frames share semantic coherence. We define a visual event as a sequence of images with consistent visual-semantic content and uninterrupted timestamps and we adopt the concept of visual events as a fundamental processing unit for analyzing continuous image sequences. 

\textbf{Temporal Expansion Method for Candidate Set Optimization.} Given the temporal continuity of lifelog data (evidenced by timestamp-based image naming conventions), we propose a temporal expansion method for candidate set optimization at the initial stage. For two temporally adjacent images, we use their CLIP-based feature measure similarity to distinguish whether they belong to different events for event slicing. 
The retrieval process based on events can be divided into three phases: First, CLIP-generated embeddings of query texts and events (the average of the included image features) are compared via cosine similarity to obtain an initial Top 100 candidate event set per query. 
Second, temporal distribution analysis is performed on these candidates by aggregating their capture times (hour-level precision) to identify peak concentration intervals. 
Finally, leveraging the sequential naming structure, we expand the candidate set by including 80 preceding and 80 subsequent images centered on the first image in the peak interval. 

Compared to conventional similarity-based retrieval, the temporal expansion strategy demonstrates distinct advantages for lifelog datasets with pronounced sequential characteristics. However, the strategy has inherent limitations. Primarily, its core mechanism relies heavily on standardized timestamp naming conventions. When the dataset lacks naming based on time specification, the system fails to accurately locate chronologically adjacent images, rendering the strategy ineffective. Secondly, the current implementation exclusively expands the most densely clustered time segments. 
This approach may overemphasize signals from dominant segments while neglecting potentially relevant information in other periods.

\textbf{Event-based Multi-Round Candidate Set Expansion.} To mitigate the weakness of relying on textual representations alone in retrieval, we propose an event-based multi-round candidate set expansion strategy. This strategy also considers the division of images in the form of events. The process is as follows: first, based on the query and event representations generated by CLIP, we got the top 100 most similar events using cosine similarity;
Second, We compute the cosine similarity between each image in the top 100 events and textual query.
Finally, the features corresponding to the top 5 images and the textual query representation are summed to obtain the new query representation.

We extend the strategy to multi-round expansion to avoid missing images with lower similarity and not included in the candidate set when the initial candidate set is obtained. Through this multi-round candidate set expansion strategy, we significantly improve the Recall of initial retrieval results and provide a richer candidate pool for subsequent fine filtering based on a multimodal large language model (MLLM).

\subsection{MLLM-based Posterior Filtering}\label{mllm_filter}
To address data heterogeneity — a core characteristic of lifelogs where data spans visual and textual — we employ Multimodal Large Language Models (MLLMs) for cross-modal alignment. Traditional cross-modal retrieval systems often face the semantic gap when dealing with complex queries, especially when there is an implicit association between query text and visual content or complex inference is required. Multimodal Large Language Models enhance cross-modal retrieval by integrating and interpreting diverse data modalities through multi-modal semantic alignment, enabling natural language queries for complex temporal and contextual patterns. In order to improve the accuracy of the experimental results and make good use of the expanded candidate set, we use Qwen2-VL~\cite{wang2024qwen2} to filter and rerank the candidate sets obtained in the previous stage. It mainly uses Qwen2-VL's fine-grained understanding of text and image fusion to judge the correlation of images in the candidate set, so as to further improve the accuracy of retrieval results.
In the posterior filtering phase, we also use the location information in the metadata. If a geographic location is explicitly specified in the location field of the metadata, we supplement the Qwen2-VL filtered prompt by appending the clause "determine if the photo was taken at the {location}". 
\section{Experiments}
We conducted retrieval for relevant images across 26 topics (13 ad-hoc topics and 13 known-item topics) in the LSAT task, submitting results seven times in total. However, since the second and seventh submissions duplicated previous ones, there were ultimately five valid submissions.
\subsection{Experimental Setup}
We made 5 valid submissions to 26 topic queries, with the methods employed in each submission detailed in the table below, using MAP, Precision at 10 and 100 (P@10 and P@100), Recall@10 (R@10) and nDCG@10 as the evaluation metric.

\begin{table}[h]
\renewcommand{\arraystretch}{1.2} 
\caption{Methods in Each Submission.}
\label{tab:method}
\begin{tabular}{lp{6cm}c}
\toprule
LSAT01 &Query Rewriting  \\
LSAT03 &Query Rewriting and MLLM-based Posterior Filtering\\
LSAT04 &Query Rewriting, MLLM-based Posterior Filtering, and Temporal-based Candidate Expansion\\
LSAT05 &Query Rewriting, MLLM-based Posterior Filtering, and Single-Round Event-based Expansion\\
LSAT06 &Query Rewriting, MLLM-based Posterior Filtering, and Multi-Round Event-based Expansion \\
\bottomrule
\end{tabular}
\end{table}

\subsection{Results}
Table~\ref{tab:results} presents the performance of our automated process on the NTCIR-18 Lifelog-6 topics. We conducted five valid submissions (excluding duplicate LSAT02/LSAT07). Through five iterations, it demonstrates progressive performance improvements. The baseline LSAT01, employing only query rewriting and CLIP-based similarity ranking, achieved a foundational mAP@100 of 0.1492. Building upon this, LSAT03 incorporated Qwen2-VL multimodal filtering, which enhanced result reliability through cross-modal consistency verification, elevating mAP@100 by 27.6\% to 0.1905 and confirming MLLMs' effectiveness in noise suppression.

LSAT04 introduced temporal-based candidate expansion, showing promise for sequential events but constrained by timestamp dependencies. Subsequent versions transitioned to more flexible event-based approaches: LSAT05 adopted single-round event-based expansion using high-similarity same-day images with metadata integration for scene validation, improving P@10 by 9.53\% and Recall@10 by 21.5\% for queries. The final LSAT06 implemented the multiround iterative event expansion based on LSAT05, achieving a peak performance of mAP = 0.2652, P@10 = 0.3038, and Recall@10 = 0.2617. Judging from the results, the addition of each of our methods has improved the evaluation. This proves the feasibility and effectiveness of the targeted methods proposed for the three characteristics of lifelog data.

\begin{table}[h]
\centering
\caption{Performance of LifeIR Submissions}
\label{tab:results}
\begin{tabular}{lrrrrr}
\toprule
RunID & MAP & P@10 & P@100 & R@10 & nDCG@10\\
\midrule
LSAT01 & 0.1492 & 0.2423 & 0.1496 & 0.1741 & 0.3079 \\
LSAT03 & 0.1905 & 0.2769 & 0.1373 & 0.1696 & 0.3584 \\
LSAT04 & 0.1887 & 0.2731 & 0.1323 & 0.1693 & 0.3587 \\
LSAT05 & 0.2103 & 0.2654 & 0.1962 & 0.2116 & 0.3294 \\
LSAT06 & 0.2652 & 0.3038 & 0.1608 & 0.2617 & 0.4590 \\
\bottomrule
\end{tabular}
\end{table}

\subsection{Bad Case Analysis}
Through manual inspection of retrieval results for each query, we identify a subset of underperforming queries along with their corresponding results. Our analysis reveals that these queries (e.g., topics titled “Putting away the Christmas tree” or “Poster about a Castle”) primarily challenge the embedding model’s fine-grained reasoning and semantic comprehension capabilities. Taking “Putting away the Christmas tree” as an example, the Christmas tree in this context may not appear in its fully decorated form but rather in a folded or stored state, requiring the model to perform reasoning to interpret the concept of “a Christmas tree being put away.” However, CLIP, as a lightweight contrastive learning-based model, exhibits limitations in fine-grained visual-semantic understanding~\cite{kudo2003prognostic}, leading to suboptimal performance on such complex queries.

\section{Conclusions}
In conclusion, this paper presents a systematic exploration by the LifeIR team for the LSAT subtask. We propose an innovative framework based on automated retrieval algorithms, with CLIP serving as the central embedding model, and adding data processing, query rewriting, event-based candidate set augmentation, and optimize experimental results using MLLM for filtering. Experimental results demonstrate that incorporating event-based candidate expansion and MLLM posterior filtering techniques further enhances the system’s precision of retrieval outcomes. At the end of Section 4, we conduct an analysis to investigate the underlying causes of suboptimal performance on specific queries. For further development, we will continue to enhance the automatic approach by upgrading the embedding model to improve its fine-grained understanding of text and images. Furthermore, while our current event-based methodology demonstrates promising results, its conceptual framework requires further refinement. Future work will focus on developing more robust event modeling techniques to enhance system performance.

\section{ACKNOWLEDGEMENT}
This work was funded by the National Natural Science Foundation of China (NSFC) under Grants No. 62302486, the Innovation Project of ICT CAS under Grants No. E361140, the CAS Special Research Assistant Funding Project, the project under Grants No. JCKY2022130C039, the Strategic Priority Research Program of the CAS under Grants No. XDB0680102, and the NSFC Grant No. 62441229.

\bibliographystyle{ACM-Reference-Format}
\bibliography{ntcirsample}

\end{document}